\documentclass[twocolumn,showpacs,amsmath,amssymb,pra]{revtex4}

\usepackage{graphicx}   	% Include figure files
\usepackage{dcolumn}    	% Align table columns on decimal point
\usepackage{bm}         	% bold math
\newcommand{\rd}{{\rm d}}
\newcommand{\ri}{{\rm i}}
\newcommand{\re}{{\rm e}}
\newcommand{\kc}{k_{\rm c}}
\newcommand{\kL}{k_{\rm L}}
\newcommand{\Er}{E_{\rm r}}
\newcommand{\Fac}{F_{\rm ac}}
\newcommand{\Kmax}{K_0^{\rm max}}
\newcommand{\TP}{T_{\rm P}}
\begin{document}

\title[Wave packet manipulation]
      {Controlled wave-packet manipulation with driven optical lattices}

\author{Stephan Arlinghaus}

\author{Martin Holthaus}

\affiliation{Institut f\"ur Physik, Carl von Ossietzky Universit\"at,
	D-26111 Oldenburg, Germany}

\date{December 14, 2011}

\begin{abstract}
Motivated by recent experimental progress achieved with ultracold atoms in kilohertz-driven optical lattices, we provide a theoretical discussion of mechanisms governing the response of a particle in a cosine lattice potential to strong forcing pulses with smooth envelope. Such pulses effectuate adiabatic motion of a wave packet's momentum distribution on quasienergy surfaces created by spatiotemporal Bloch waves. Deviations from adiabaticity can then deliberately be exploited for exerting coherent control and for reaching target states which may not be accessible by other means. As one particular example, we consider an analog of the $\pi$ pulses known from optical resonance. We also suggest adapting further techniques previously developed for controlling atomic and molecular dynamics by laser pulses to the coherent control of matter waves in shaken optical lattices.  
\end{abstract}


% 67.85.Hj 	Bose-Einstein condensates in optical potentials
% 42.50.Hz 	Strong-field excitation of optical transitions in quantum systems; multiphoton processes; dynamic Stark shift
% 32.80.Qk 	Coherent control of atomic interactions with photons 
% 03.75.Lm 	Tunneling, Josephson effect, Bose-Einstein condensates in periodic potentials, solitons, vortices, and topological excitations 

%\keywords{Optical lattices, spatiotemporal Bloch waves, Floquet theory, 
%		adiabatic response, pulse shaping, coherent control}

\maketitle

%%%%%%%%%%%%%%%%%%%%%%%%%%%%%%%%%%%%%%%%%%%%%%%%%%%%%%%%%%%%%%%%%%%%%%%%%%%%%%%%

\section{Introduction}

Boosted by the seminal observation of the quantum phase transition from a superfluid to a Mott insulator in a gas of ultracold $^{87}$Rb atoms trapped by an optical lattice potential~\cite{GreinerEtAl02}, the experimental and theoretical study of ultracold atoms in optical lattices has matured into a major area of contemporary research~\cite{MorschOberthaler06,LewensteinEtAl07,BlochEtAl08}. To a large extent, this field is driven by the promise of simulating complex condensed-matter systems and obtaining novel insight into phenomena which hitherto are not understood, such as high-temperature superconductivity.

At present, evidence is accumulating which suggests that this field is developing a new branch, aiming at the coherent control of mesoscopic matter waves in optical lattices through the application of time-periodic forces, with driving frequencies in the lower kilohertz regime. While it had been pointed out earlier that a metal-insulator--like transition undergone by ultracold atoms in quasiperiodic optical lattices should be controllable by adjusting the amplitude of a sinusoidal drive~\cite{DreseHolthaus97}, experimental work in this direction increased pace only in 2007, with the clear-cut observation of forcing-induced dynamical suppression of tunneling, and even reversal of the sign of the tunneling matrix element, by Arimondo, Morsch, and co-workers~\cite{LignierEtAl07,EckardtEtAl09}. This group also has documented an analog of photon-assisted tunneling with Bose-Einstein condensates in shaken optical lattices~\cite{SiasEtAl08}, and has verified that the superfluid--to--Mott-insulator transition can be coherently controlled by suitably ``dressing'' a matter wave in an optical lattice~\cite{ZenesiniEtAl09}, taking up a theoretical proposal by Eckardt {\em et al.\/}~\cite{EckardtEtAl05}. The very same principle underlying this form of coherent control has quite recently been exploited successfully for emulating frustrated magnetism in driven triangular optical lattices~\cite{StruckEtAl11}, which may well be regarded as a guiding landmark example of quantum simulation. Moreover, there now exist first experimental results demonstrating active control of correlated tunneling in ac-driven optical lattices~\cite{ChenEtAl11}.  

These experimental advances concerning time-periodically-driven matter waves are accompanied by growing theoretical efforts. For example, Kudo {\em et al.\/} have investigated the possibility of driving-induced control of bound-pair transport~\cite{KudoEtAl09}, while Tokuno and Giamarchi have studied a kind of spectroscopy for cold atoms in periodically phase-modulated optical lattices~\cite{TokunoGiamarchi11}. Moreover, Tsuji {\em et al.\/} have pointed out that ac forcing may even change the interparticle interaction from repulsive to attractive, possibly allowing one to simulate an effectively attractive Hubbard model with a temperature below the superconducting transition temperature~\cite{TsujiEtAl11}. So far, all these considerations merely involve strict ac forcing with a constant amplitude. In analogy to the physics of atoms and molecules interacting with laser pulses, here we suggest that many more control options should become available when ultracold atoms in optical lattices are subjected to forcing {\em pulses\/} with a deliberately shaped envelope.  
 
Such attempts to gain coherent control over mesoscopic matter waves call for a systematic theory of the response of ultracold many-body systems to nonperturbatively strong external forcing. Although one may obtain some insight from drastically simplified model systems~\cite{Holthaus01}, and substantial progress is being made now with the help of advanced numerical schemes~\cite{PolettiKollath11}, this goal still is far from accomplishment. In this situation, an intermediate step suggests itself: In experiments with sufficiently dilute Bose-Einstein condensates in optical lattices, or with condensates for which the interparticle $s$-wave scattering length has been tuned close to zero by means of a Feshbach resonance, one may ignore interaction effects altogether, and can observe typical single-particle phenomena, such as ordinary or ``super'' Bloch oscillations, with condensates~\cite{IvanovEtAl08,GustavssonEtAl08,AlbertiEtAl09,HallerEtAl10}. Thus, it seems advisable to undertake a comprehensive theoretical study of the possibilities of coherent control of single-particle dynamics in forced optical lattices. The results of such a study can then immediately be tested in ``interaction-free'' condensate experiments and may help one to disentangle genuine many-body effects at a later stage. 

This is the step we are going to take in the present paper. Building on our previous work~\cite{ArlinghausHolthaus10}, here we provide a detailed picture of basic mechanisms which imply single-particle state control: We first demonstrate in Sec.~\ref{sec:S_2} the feasibility of adiabatic transport of momentum distributions on quasienergy surfaces corresponding to time-periodically-forced optical lattices. This option is opened up by the existence of a basis of spatiotemporal Bloch waves, that is, of Bloch-like states which embody both the spatial periodicity of the lattice and the temporal periodicity of a driving force on equal footing; such states constitute the foundation of our analysis~\cite{ArlinghausHolthaus11}. We then establish in Sec.~\ref{sec:S_3} an analog of the $\pi$ pulses known from the theory of  optical resonance~\cite{AllenEberly75} and outline how to utilize avoided crossings of quasienergy surfaces for ``cutting out'' parts of an initially given momentum distribution. Section~\ref{sec:S_4} briefly addresses effects connected to the phase of the driving force. Taken together, our findings indicate that there is a high potential for transferring well-established methods currently used for manipulating and controlling atoms and molecules by specifically designed laser pulses~\cite{JudsonRabitz92,Rice92,AssionEtAl98,ShapiroBrumer03,Baumert11} to the newly emerging field of manipulating and coherently controlling mesoscopic matter waves in optical lattices by specifically tailored forces; this prospect is put forward in our conclusions.

\section{Adiabatic transport of momentum distributions}
\label{sec:S_2}    

The starting point of our considerations is a single particle of mass $m$ moving in a one-dimensional optical lattice potential~\cite{MorschOberthaler06,LewensteinEtAl07,BlochEtAl08}  
\begin{equation}
  	V(x) = \frac{V_0}{2}\cos\left(2\kL x\right) \; ,
\label{eq:lattice}
\end{equation}
where the lattice depth $V_0$ is proportional to the intensity of the laser radiation generating the lattice, and $\kL$ denotes the corresponding wave number. Thus, the potential is periodic with lattice constant $a = \pi/\kL$, so that $V(x) = V(x+a)$. Moreover, the particle is subjected to a spatially homogeneous inertial force $F(t)$, which can be applied by accelerating the lattice in the laboratory frame~\cite{Graham92}. After transforming to a frame of reference co-moving with the lattice, the Hamiltonian of the system is given by
\begin{equation}
  	\widetilde H(x, t) = \frac{p^2}{2m} + V(x) - F(t)x \; .
\label{eq:tildeH}
\end{equation}
If we denote the solution to the Schr\"odinger equation pertaining to Eq.~\eqref{eq:tildeH} by $\widetilde\psi(x, t)$ and perform the unitary transformation 
\begin{equation}
  	\psi(x, t) = \exp\!\left(-\frac{\ri}{\hbar} x \int_0^t \! \rd \tau \, F(\tau) \right) \widetilde\psi(x, t) \; ,
\label{eq:U}
\end{equation}
the transformed functions $\psi(x, t)$ obey a Schr\"odinger equation with the new Hamiltonian
\begin{equation}
  	H(x, t) = 
	\frac{1}{2m}\left(p + \int_0^t \! \rd \tau \, F(\tau) \right)^2 + V(x) \; .
\label{eq:H}
\end{equation}
The traditional solid-state approach to monitoring the wave-packet dynamics now is as follows: The unforced lattice possesses improper energy eigenstates $\chi_{n,k}(x)$ which have the form of Bloch waves~\cite{Callaway76,AshcroftMermin76,Kittel87}, that is, of plane waves which are modulated by lattice-periodic functions $v_{n,k}(x) = v_{n,k}(x+a)$, so that
\begin{equation}
	\chi_{n,k}(x) = \re^{\ri kx} v_{n,k}(x) \; ;
\label{eq:BW}
\end{equation}
these waves solve the time-independent Schr\"odinger equation
\begin{equation}
  	\left( \frac{p^2}{2m} + V(x) \right) \chi_{n,k}(x) = E_n(k) \chi_{n,k}(x) \; .
\label{eq:TIS}	 
\end{equation}
Here $n$ is a band index and $k$ a wave number, so that $E_n(k)$ is the energy dispersion relation of the $n$th Bloch band. Owing to the periodicity of the lattice, the wave numbers can be restricted to the first quasimomentum Brillouin zone $\mathcal{B} = [-\pi/a, +\pi/a[$. In addition, we require the normalization
\begin{equation}
	\int_{-\infty}^{+\infty} \! \rd x \, \chi_{n',k'}^*(x) \chi_{n,k}(x) = \frac{2\pi}{a} \delta_{n, n'} \delta(k - k') \; .
\label{eq:NormB}
\end{equation}
When an arbitrary given wave packet $\psi(x, t)$ is expanded with respect to the Bloch basis in the form
\begin{equation}
  	\psi(x, t) = \sum_n \sqrt{\frac{a}{2\pi}} \int_{\mathcal{B}} \rd k\, g_n^{\rm B}(k,t) \chi_{n, k}(x) \; ,
\label{eq:CMR}	
\end{equation}
this convention~\eqref{eq:NormB} makes sure that the momentum distributions $|g_n^{\rm B}(k,t)|^2$ are normalized according to
\begin{equation}
	\sum_n \int_{\mathcal{B}} \rd k\, |g_n^{\rm B}(k,t)|^2 = 1 \; .
\end{equation}
In solid-state physics, the expansion~\eqref{eq:CMR} is known as the crystal-momentum representation of the wave packet $\psi(x, t)$.   

When there is no external forcing, $F(t) \equiv 0$, the time dependence of the expansion coefficients $g_n^{\rm B}(k,t)$ in Eq.~\eqref{eq:CMR} simply reads
\begin{equation} 
  	g_n^{\rm B}(k,t) = g_n^{\rm B}(k,0) \re^{-\ri E_n(k)t/\hbar} \; .
\end{equation}\
For studying the dynamics under the action of a force $F(t)$, let us at this point assume that the wave packet initially occupies only one band with a particular index~$n$, and that the force remains so weak that it does not induce substantial interband transitions. Then Bloch's acceleration theorem~\cite{Callaway76,AshcroftMermin76,Kittel87} comes into play: The packet's center wave number in $k$~space, given by the first moment   
\begin{equation}
	k_{\rm c}(t) = \int_{\mathcal{B}} \rd k \, k |g_n^{\rm B}(k,t)|^2 \; , 
\end{equation}
then evolves according to the semiclassical law
\begin{equation}
  	\hbar\dot{k}_{\rm c}(t) = F(t) \; .
\label{eq:ACT}	
\end{equation}
For example, a constant force leads to a linearly increasing $k_{\rm c}(t)$, which, in its turn, gives rise to Bloch oscillations in real space.

Although this time-honored approach has many virtues, for our purposes it is advantageous to look at the wave-packet dynamics from a different angle. In view of the goal to exert coherent control on the lattice atom, it is quite natural to specifically consider sinusoidal forces $F(t) = F_{\rm ac}\sin(\omega t)$ in the first place, since then the Hamiltonian~\eqref{eq:tildeH} is of the familiar form which also describes a charged particle in a monochromatic classical radiation field within the dipole approximation. The most conspicuous difference concerns the frequencies: Typical frequencies for driving optical lattices~\cite{LignierEtAl07,EckardtEtAl09,SiasEtAl08,ZenesiniEtAl09,EckardtEtAl05,StruckEtAl11,ChenEtAl11} fall into the lower kilohertz regime, about 11 orders of magnitude lower than optical frequencies. In the following, we merely require that the force be periodic in time with period~$T$, so that $F(t) = F(t+T)$, and we assume that its one-cycle average vanishes, so that
\begin{equation}
  	\frac{1}{T} \int_0^T \! \rd t \, F(t) = 0 \; . 
\end{equation}
With these specifications, the transformed Hamiltonian~\eqref{eq:H} is periodic in space as well as in time, $H(x,t) = H(x+a,t) = H(x,t+T)$. While the ordinary Bloch waves~\eqref{eq:BW} account for the spatial periodicity only, the mathematical Floquet theorem governing the structure of solutions to differential equations with periodic coefficients~\cite{Floquet83,Eastham73,Kuchment93} can now be invoked to simultaneously incorporate {\em both\/} the spatial {\em and\/} the temporal periodicity, resulting in a set of solutions to the time-dependent Schr\"odinger equation of the suggestive form  
\begin{equation}
  	\psi_{n, k}(x, t) = \exp\left[\ri kx - \ri\varepsilon_n(k)t/\hbar\right] u_{n, k}(x, t)
\label{eq:spatiotemporalBW}
\end{equation}
with biperiodic functions $u_{n,k}(x,t)$ which reflect the two translational symmetries, $u_{n,k}(x,t) = u_{n,k}(x+a,t) = u_{n,k}(x,t+T)$. We refer to these solutions~\eqref{eq:spatiotemporalBW} as {\em spatio\-temporal Bloch waves\/}~\cite{ArlinghausHolthaus11}. The quantities $\varepsilon_n(k)$ determining the linear growth of the phase factors with time are commonly known as quasienergies~\cite{Zeldovich66,ChuTelnov04}. They are obtained by solving the eigenvalue problem
\begin{equation}
  	\left(H(x, t) - \ri\hbar\frac{\partial}{\partial t}\right) \varphi_{n, k}(x, t) = \varepsilon_n(k)\varphi_{n, k}(x, t) \; , 
\label{eq:EVE}
\end{equation}
where the functions $\varphi_{n, k}(x, t) = \exp\left(\ri kx\right)u_{n, k}(x, t)$ denote the spatial parts of the spatiotemporal Bloch waves~\eqref{eq:spatiotemporalBW}. Fully in accordance with our rationale, this eigenvalue problem~\eqref{eq:EVE} is posed in an extended Hilbert space which puts position~$x$ and time~$t$ on equal footing~\cite{Sambe73}. Because of the periodicity of $H(x,t)$ in time, the eigenvalues $\varepsilon_n(k)$ are defined up to an integer multiple of $\hbar\omega$, with $\omega = 2\pi/T$, which means that there also is a Brillouin-zone scheme for the quasienergies, with the fundamental zone $\mathcal{Q} = [-\hbar\omega/2, +\hbar\omega/2[$, in analogy to the fundamental quasimomentum zone $\mathcal{B} = [-\pi/a, +\pi/a[$. Evidently, Eq.~\eqref{eq:EVE} now takes the place of the traditional eigenvalue equation~(\ref{eq:TIS}), and the spatiotemporal Bloch waves~\eqref{eq:spatiotemporalBW} replace the ordinary Bloch waves~\eqref{eq:BW}. Consequently, we abandon the standard crystal-momentum representation~\eqref{eq:CMR} and instead perform expansions of given wave packets $\psi(x,t)$ in this new basis: Fixing, in  analogy to the previous Eq.~\eqref{eq:NormB}, the normalization     
\begin{equation}
  	\int_{-\infty}^{\infty} \rd x\, 
	\varphi_{n', k'}^{*}(x, t) \varphi_{n, k}(x, t) = \frac{2\pi}{a}\delta_{n, n'} \delta\left(k - k'\right) \; ,
\label{eq:norm}
\end{equation}
we thus arrive at the Floquet representation~\cite{ArlinghausHolthaus11}
\begin{equation}
  	\psi(x, t) = \sum_n \sqrt{\frac{a}{2\pi}} \int_{\mathcal{B}} \rd k \, g_n(k,t) \varphi_{n, k}(x, t) \; .
\label{eq:FLR}
\end{equation}
When the amplitude of the driving force goes to zero, the functions $\varphi_{n, k}(x, t)$ reduce to the Bloch waves~\eqref{eq:BW}, and the quasienergies $\varepsilon_n(k)$ approach the energies $E_n(k)$, modulo $\hbar\omega$. Therefore, in this limit the Bloch expansion~\eqref{eq:CMR} coincides with the Floquet expansion~\eqref{eq:FLR}. However, in the presence of a strictly $T$-periodic force, such as $F(t) = \Fac\sin(\omega t)$, we now have two different pictures of the same wave-packet dynamics: Within the crystal-momentum approach, a single-band wave packet is described by a momentum distribution $|g_n^{\rm B}(k,t)|^2$; the center of this distribution moves in $k$~space according to the acceleration theorem~\eqref{eq:ACT}. In contrast, within the Floquet picture one merely has
\begin{equation} 
  	g_n(k,t) = g_n(k,0) \re^{-\ri \varepsilon_n(k)t/\hbar}
	\label{eq:constcoeff} \; ,
\end{equation}
so that the Floquet distribution $|g_n(k,t)|^2 = |g_n(k,0)|^2$ does not move at all, but stays perfectly constant in time, with the response to the oscillating force already being incorporated into the basis states~\eqref{eq:spatiotemporalBW}.

Clearly, both approaches are mathematically equivalent. But it is the second one which allows us to make further contact with advanced techniques developed for studying the interaction of atoms and molecules with laser radiation, and for developing schemes for coherent control of ultracold atoms in driven optical lattices. 

Such schemes naturally will involve {\em pulses\/} of driving forces, that is, non-periodic forcing. As a simple case, we may consider pulses of the form 
\begin{equation}
	F(t) = \Fac^{\max} s(t) \sin(\omega t) \; ,
\label{eq:PUL}	
\end{equation}
where the dimensionless shape function $s(t)$ vanishes before and after the pulse, $s(t) = 0$ for both $t < 0$ and $t > \TP$, say, and is normalized such that its maximum value is 1, implying that $\Fac^{\max}$ is the maximum amplitude encountered during the pulse.  

It is then of key importance to note that the Floquet picture is meaningful not only for perfectly time-periodic forces $F(t) = F(t+T)$, but also for situations in which one or more system parameters change slowly, that is, undergo only minor variations during one cycle~$T$~\cite{DreseHolthaus99}. This is the case, for instance, if the pulse $F(t)$ is equipped with a ``slowly'' varying envelope $s(t)$. Then one considers not only one single eigenvalue problem~\eqref{eq:EVE} corresponding to one particular amplitude $\Fac$, but rather the family of all such eigenvalue problems with $0 \le \Fac \le \Fac^{\max}$. This gives a basis of spatiotemporal Bloch waves for each instantaneous value of the amplitude; taken together, these bases serve as a ``moving frame of reference'' with respect to which the wave packet can evolve adiabatically: Under pulse conditions enabling adiabaticity, the Floquet momentum distributions $|g_n(k,t)|^2$ remain almost constant in time, provided the expansion~\eqref{eq:FLR} refers at each moment to that basis of spatiotemporal Bloch waves which is obtained by fixing the slowly varying amplitude at its momentary value. Pictorially speaking, the instantaneous eigenvalues $\varepsilon_n^{\Fac}(k)$, considered as functions of wave number~$k$ and driving amplitude~$\Fac$, form quasienergy surfaces on which the momentum distribution can move almost without change of shape in response to smooth variations of the envelope $s(t)$.    

To see what this means in practice, we consider a cosine lattice~\eqref{eq:lattice} with depth $V_0 = 5.7\,\Er$, where $\Er = \hbar^2\kL^2/(2m)$ denotes the single-photon recoil energy~\cite{MorschOberthaler06,LewensteinEtAl07,BlochEtAl08}; this depth is routinely being realized in current experiments~\cite{LignierEtAl07,EckardtEtAl09,SiasEtAl08,ZenesiniEtAl09,StruckEtAl11,ChenEtAl11}. The width of the lowest Bloch band then amounts to $E_1(\kL) - E_1(0) = 0.220\,\Er$, while the lowest band gap is $E_2(\kL) - E_1(\kL) = 2.763\,\Er$. The maximum separation of the lowest two bands, encountered in the Brillouin zone center, figures as $E_2(0) - E_1(0) = 4.690\,\Er$~\cite{ArlinghausHolthaus10}. We then take a wave packet prepared at time $t = 0$ in the lowest Bloch band $n = 1$,   
\begin{equation}
  	\psi(x, 0) = \sqrt{\frac{a}{2\pi}} \int_{\mathcal{B}} \rd k \, g_1^{\rm B}(k, 0) \chi_{1,k}(x) \; ,
\label{eq:INI}
\end{equation}
with a Gaussian initial momentum distribution 
\begin{equation}
  	g_1^{\rm B}(k, 0) = \left(\sqrt{\pi}\Delta k\right)^{-1/2} \exp\left(-\frac{k^2}{2\left(\Delta k\right)^2}\right)
\label{eq:IMD}
\end{equation}
centered around $\kc(0)/\kL = 0$ with width $\Delta k / \kL = 0.1$. This wave packet is subjected to pulses~\eqref{eq:PUL} of length $\TP$ with a squared-sine envelope,
\begin{equation}
  	s(t) = \sin^2\left(\pi \frac{t}{\TP}\right) \quad ; \quad 0 \le t \le \TP \; . 
\label{eq:env}
\end{equation}
The frequency selected for the model calculation discussed in the following is $\omega = 1.640\,\Er/\hbar$, well below the lowest band gap, while the maximum scaled driving amplitude 
\begin{equation}
	\Kmax = \frac{\Fac^{\rm max}a}{\hbar\omega}
\label{eq:K0max}
\end{equation}
is set to $\Kmax = 0.8$. The pulse length is fixed at $\TP = 10\,T$, so that the peak driving strength is reached within no more than five cycles. We then compute, on the one hand, the momentum distributions $|g_1^{\rm B}(k, t)|^2$ in the crystal-momentum representation, and also perform the Floquet expansions
\begin{equation}
  	\psi(x, t) = \sum_n \sqrt{\frac{a}{2\pi}} \int_{\mathcal{B}} \rd k \, g_n(k,t) \varphi_{n, k}^{\Fac(t)}(x, t) 
\end{equation}
with respect to the instantaneous solutions $\varphi_{n, k}^{\Fac}(x, t)$ to the quasienergy equation~\eqref{eq:EVE} in order to obtain the corresponding Floquet distributions $|g_1(k, t)|^2$, on the other.

\begin{figure}[t]
\centering
\includegraphics[width = 0.9\linewidth]{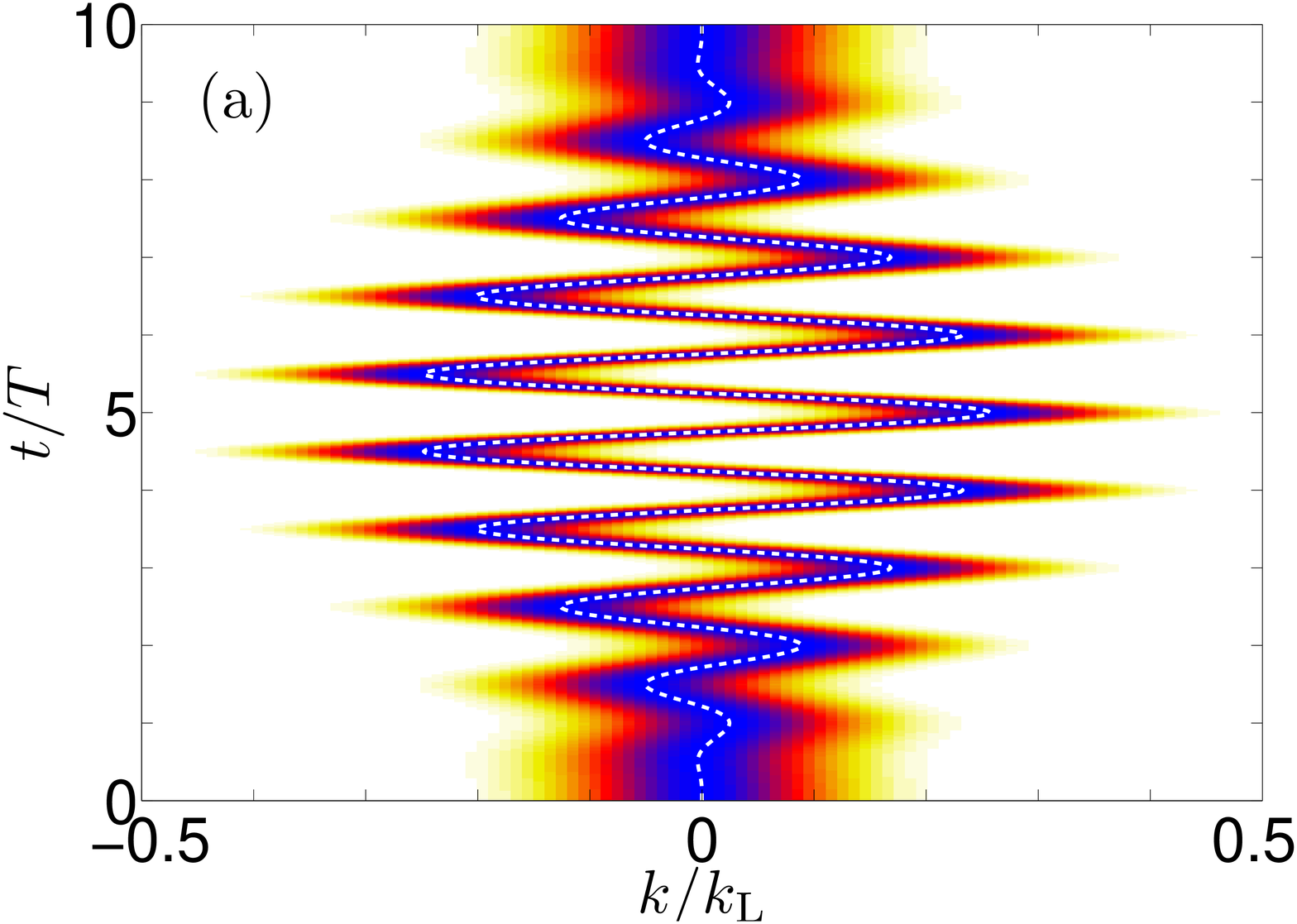}
\includegraphics[width = 0.9\linewidth]{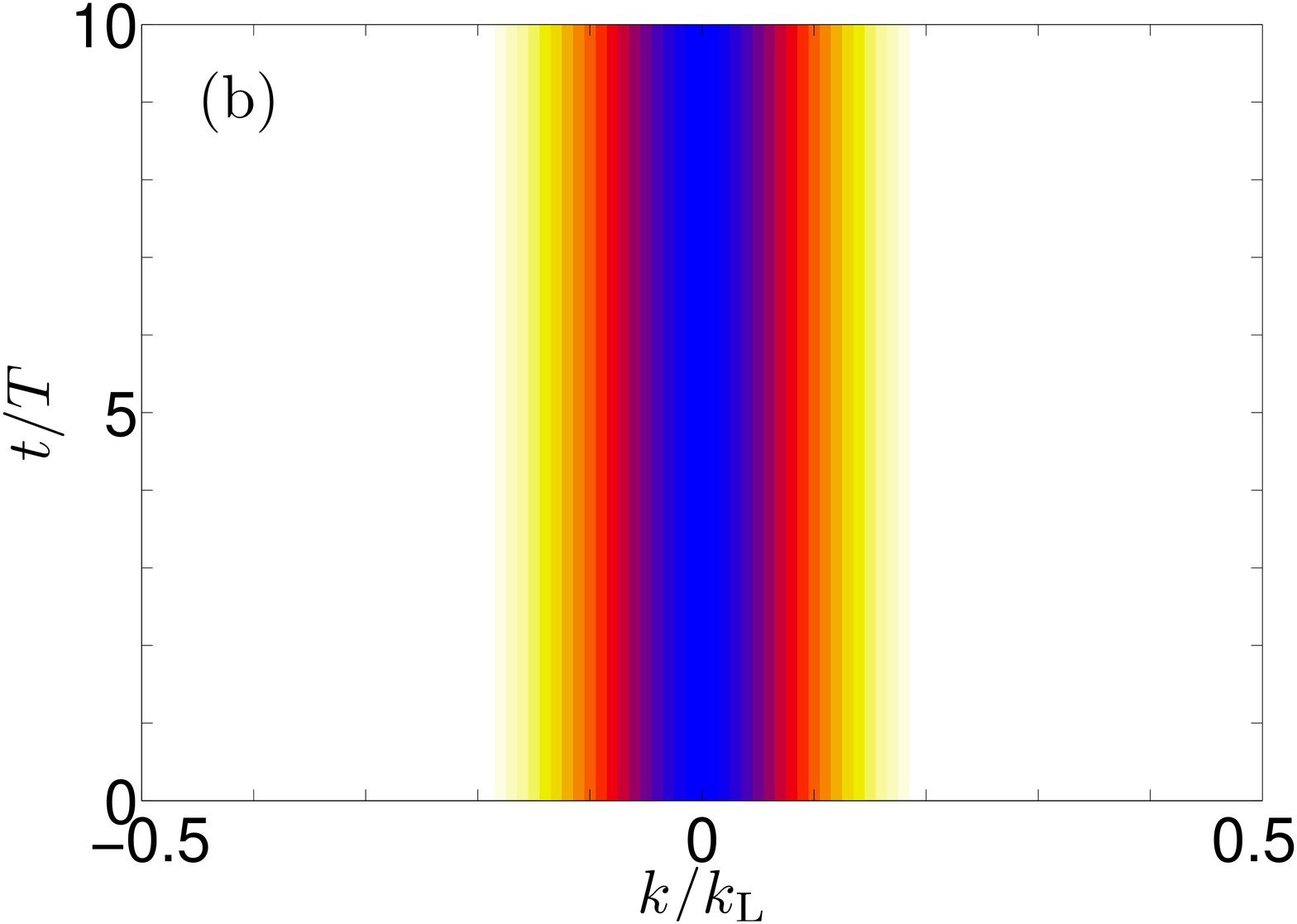}
\caption{(Color online) Response of the initial wave packet~\eqref{eq:INI} with momentum distribution~\eqref{eq:IMD} in an optical lattice with depth $V_0/\Er = 5.7$ to a short pulse~\eqref{eq:PUL} with nonresonant scaled frequency $\hbar\omega/\Er = 1.640$, maximum scaled amplitude $\Kmax = 0.8$, and pulse length  $\TP/T = 10$. (a) shows the density $|g_1^{\rm B}(k, t)|^2$ in the crystal-momentum representation. For comparison, the white-dashed line is the first moment $\kc(t)$, as predicted by the acceleration theorem~\eqref{eq:ACT}. (b) depicts the Floquet density $|g_1(k, t)|^2$, obtained by expanding the same wave packet with respect to the instantaneous spatiotemporal Bloch waves.}
\label{fig:F_1}
\end{figure}

Figure~\ref{fig:F_1} juxtaposes the results of the two approaches. In Fig.~\ref{fig:F_1}(a) we show the evolution of the crystal-momentum density $|g_1^{\rm B}(k, t)|^2$ under the pulse. Because interband transitions remain negligible for the parameters chosen, the packet's center moves in perfect accordance with Bloch's acceleration theorem~\eqref{eq:ACT}. On the other hand, Fig.~\ref{fig:F_1}(b) depicts the evolution of the Floquet density $|g_1(k, t)|^2$. This density remains practically constant, indicating almost perfect adiabatic following of $\psi(x,t)$ with respect to the spatiotemporal Bloch waves: Despite the short duration of the pulse, the initial distribution merely makes an adiabatic excursion on its quasienergy surface, returning more or less unaltered.    

\begin{figure}[t]
\centering
\includegraphics[width = 0.9\linewidth]{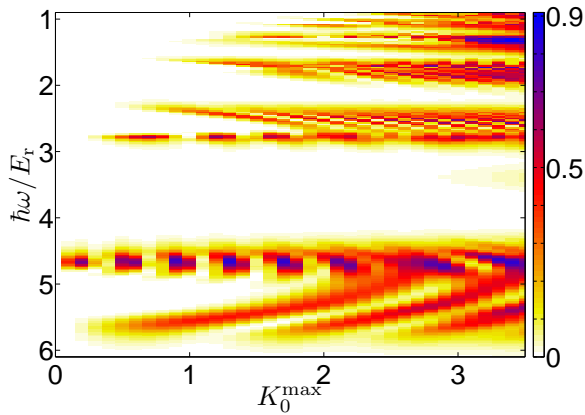}
\caption{(Color online) Final escape probabilities from the lowest Bloch band of an optical lattice with depth $V_0/\Er = 5.7$, calculated for the same initial wave packet as considered in Fig.~\ref{fig:F_1}, after pulses with squared-sine envelope~\eqref{eq:env} and length $\TP/T = 50$. Observe the window of almost adiabatic response appearing between the two-photon and the single-photon resonances.}
\label{fig:F_2}
\end{figure}

Still, Fig.~\ref{fig:F_1} is no more than a look at the same dynamics from two different viewpoints, and so far neither of these is better than the other. But now comes the crucial step: Control is exerted by utilizing interband transitions. While such transitions fall outside the scope of the semiclassical acceleration theorem, which explicitly requires a single-band setting, they can be monitored as deviations from adiabaticity, caused by near-degeneracies of quasienergy surfaces, within the Floquet approach. In order to locate the parameters for which such deviations occur, Fig.~\ref{fig:F_2} shows the final escape probability from the lowest Bloch band in the amplitude-frequency plane~\cite{ArlinghausHolthaus10}, as resulting from the same initial wave packet as constructed above after pulses with the envelope~\eqref{eq:env}, with greater length $\TP = 50\,T$. Most notably, the ``single-photon resonance'' with $\hbar\omega = E_2(0) - E_1(0)$ shows up already for quite small driving amplitudes around $\hbar\omega \approx 4.690 \, \Er$; and one observes a sequence of multiphoton-like resonances at lower frequencies. Interestingly, there also is a pronounced frequency window between the two-photon resonance and the single-photon peak which allows for adiabatic response even to fairly strong pulses with $\Kmax > 3$. This window appears to be most suitable for studying single-band phenomena associated with strong forcing, such as the driving-induced reversal of the sign of the effective hopping matrix element~\cite{LignierEtAl07,EckardtEtAl09}.

The Floquet approach now enables one to look into the transition dynamics in great detail, and thus to understand basic principles allowing one to deliberately manipulate the momentum distribution and to create certain desired target states; this will be elaborated in the following section. Before closing the present section, we would like to draw an interesting comparison: Aside from degeneracies, the adiabatic motion of a wave packet's momentum distribution on its quasienergy surface, as visualized in Fig.~\ref{fig:F_1}(b), seems to resemble the adiabatic evolution of molecular states on their Born-Oppenheimer potential energy surfaces~\cite{BornOppenheimer27}. There is, however, an important difference: In the case of cold atoms in driven optical lattices the concept of adiabatic following~\cite{BornFock28,Berry84} has to be applied to each wave number~$k$ in parallel, each one labeling a different spatiotemporal Bloch wave. Thus, here we are confronted not with adiabatic following of individual states, but rather with that of a density associated with a continuum of quasienergy eigenstates.

\section{Tailoring the momentum distribution}
\label{sec:S_3}

Non-adiabatic transitions, which prevent an initial momentum distribution from returning practically unchanged after a pulse, result from near-degeneracies of quasienergy surfaces. Two different cases have to be distinguished: Either the near-degeneracy is induced already at small driving amplitudes by selecting a resonant frequency, or it shows up only under strong nonresonant driving, when the ac Stark shift forces two surfaces into an avoided crossing~\cite{ArlinghausHolthaus10}. In this section we show that either of these scenarios can be exploited for controlling and reshaping the $k$~space distribution coherently. In all model calculations we consider an optical lattice with depth $V_0 = 5.7 \, \Er$ and start from the initial Bloch wave packet~\eqref{eq:INI} with Gaussian coefficients~\eqref{eq:IMD}, again setting $\Delta k / \kL = 0.1$.

\subsection{Resonant forcing}
\label{sec:S_3a}

\begin{figure}[t]
\centering
\includegraphics[width = 0.8\linewidth]{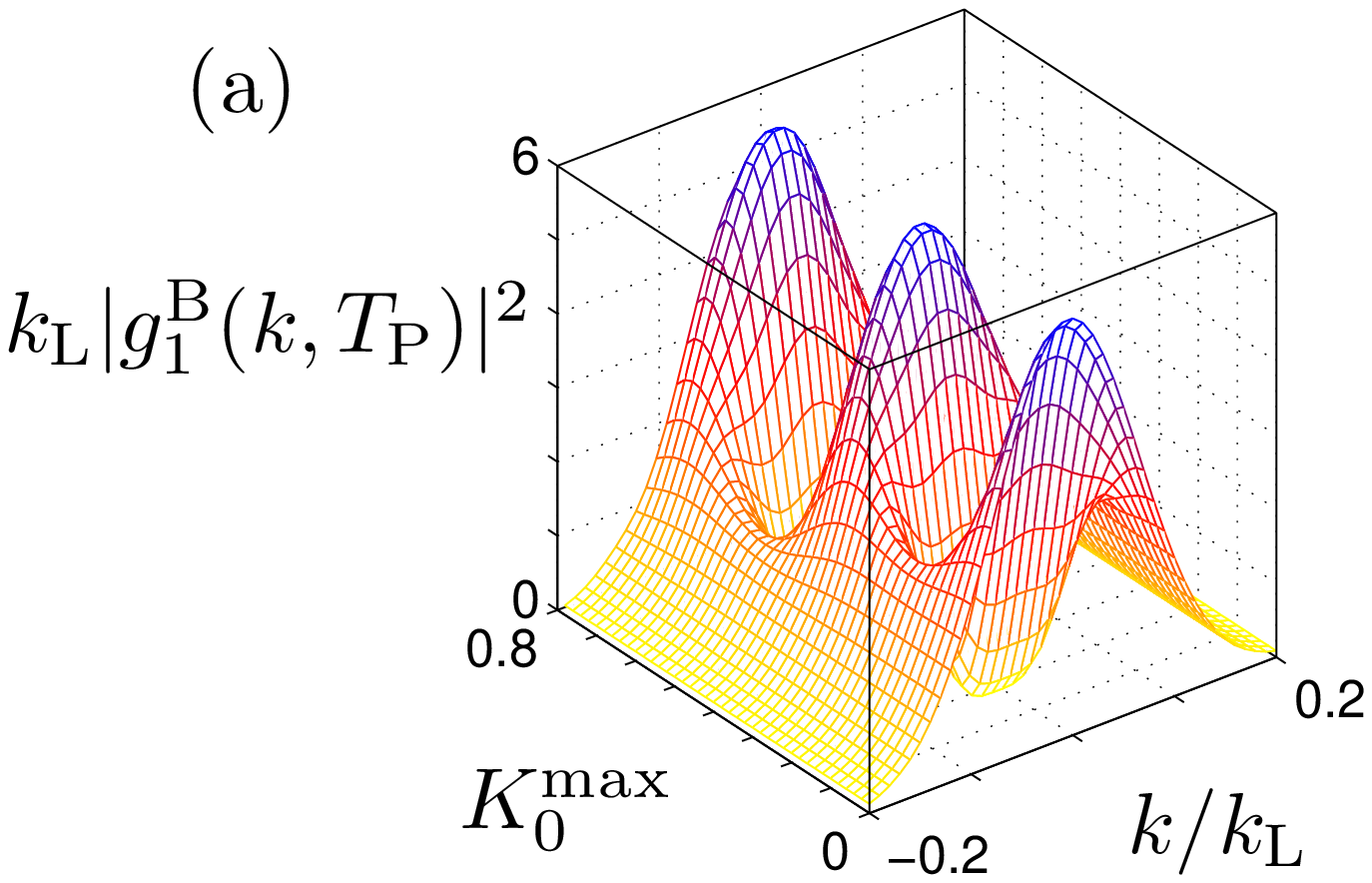}
\includegraphics[width = 0.8\linewidth]{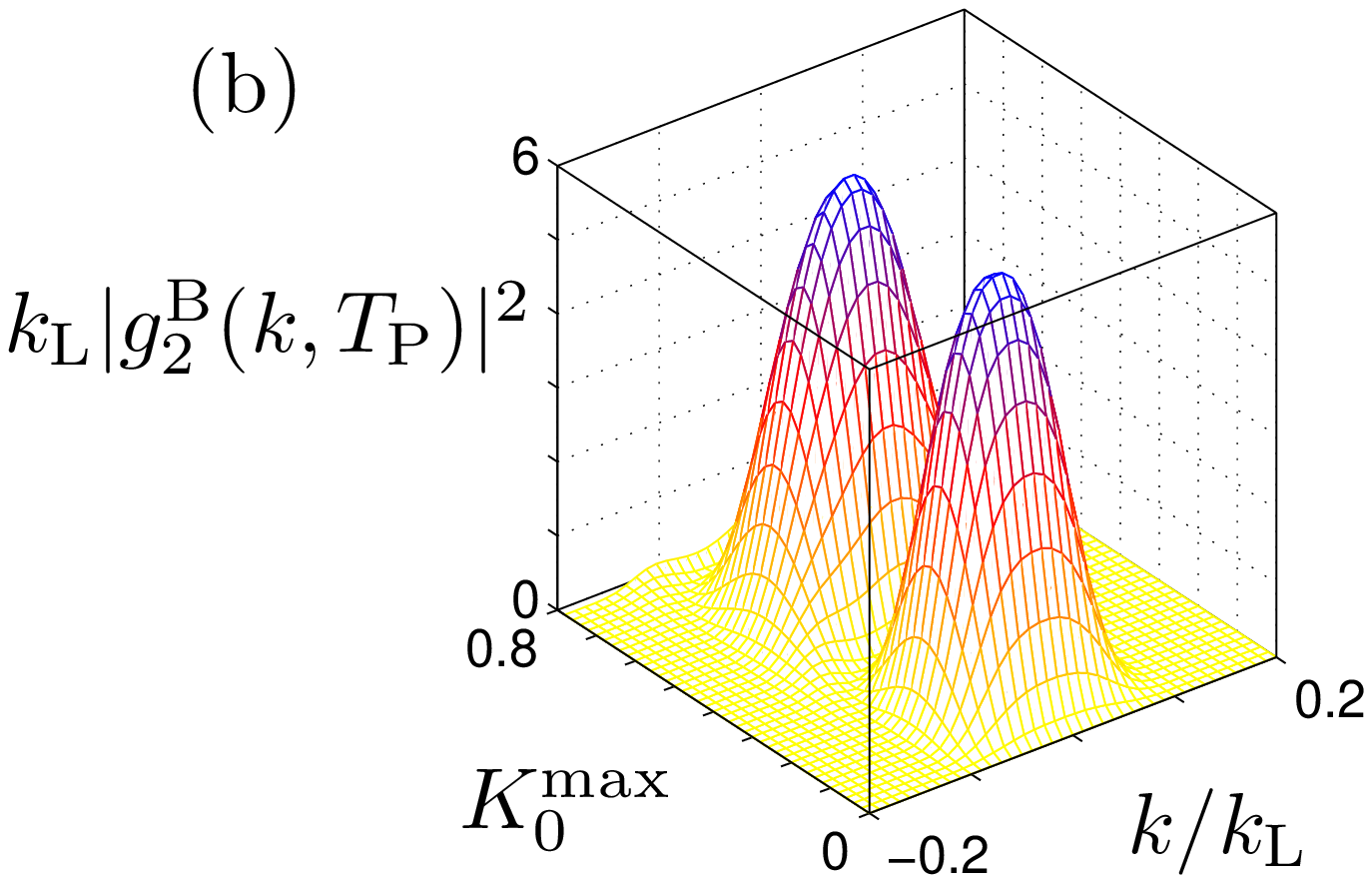}
\includegraphics[width = \linewidth]{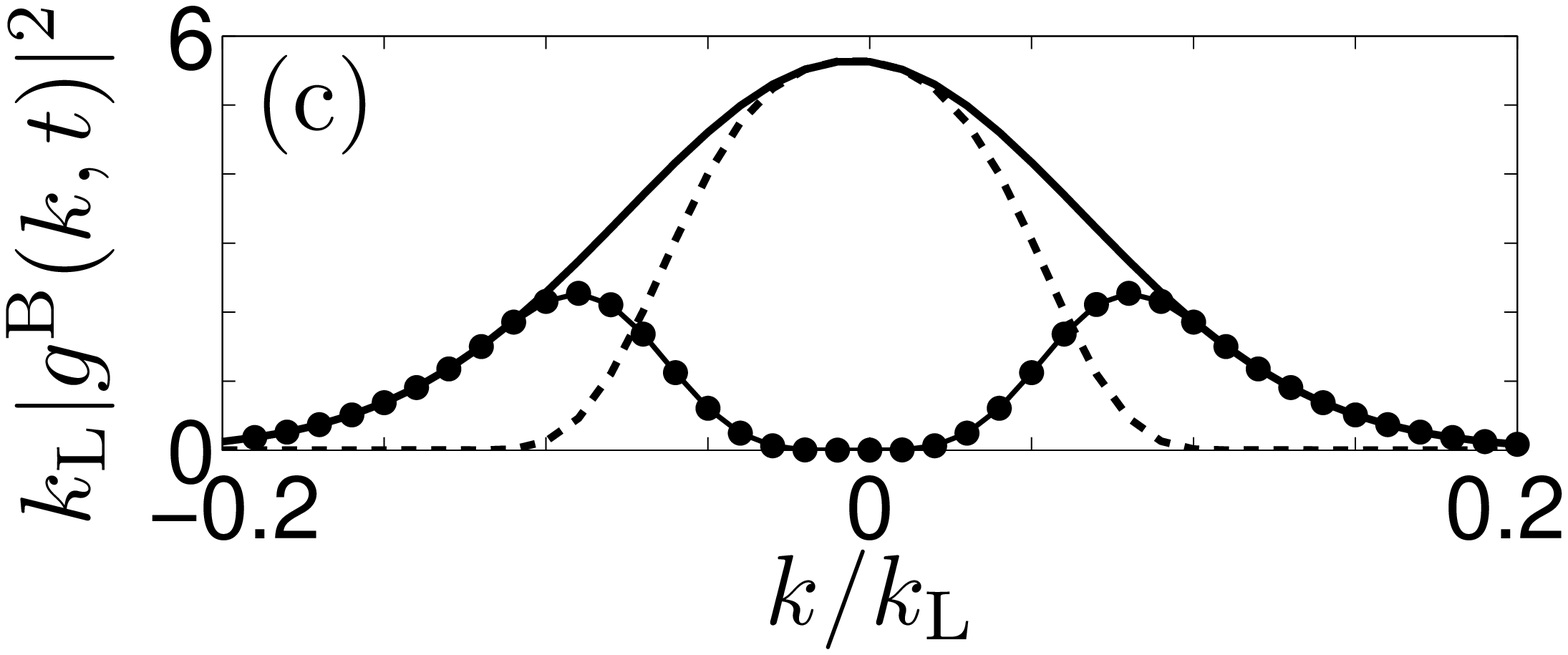}
\caption{(Color online) Final momentum distributions in the lowest (a) and in the first excited (b) Bloch band after the initial state has been exposed to pulses with squared-sine envelope~\eqref{eq:env} and length $\TP/T = 50$, with varying maximum scaled amplitudes $0 \le \Kmax \le 0.8$. Here the scaled driving frequency is $\hbar\omega/\Er = 4.690$, implying $\hbar\omega = E_2(0) - E_1(0)$, so that both bands are exactly resonant at $k / \kL = 0$. The lowest panel (c) compares the final distributions in the first (dotted) and in the second (dashed) bands to the initial distribution (full line), for $\Kmax = 0.186$. This particular situation corresponds to a ``$\pi$ pulse''.}
\label{fig:F_3}
\end{figure}

We now adjust the driving frequency such that $\hbar\omega = E_2(0) - E_1(0)$, so that the lowest two bands are coupled resonantly in the center of the quasimomentum Brillouin zone, at $k / \kL = 0$. The length of the pulses with squared-sine envelope~\eqref{eq:env} is $\TP = 50 \, T$.  Figure~\ref{fig:F_3} shows the resulting final distributions $|g_1^{\rm B}(k,\TP)|^2$ and $|g_2^{\rm B}(k,\TP)|^2$ for the lowest and for the first excited Bloch band, respectively, in dependence on the maximum driving amplitude~$\Kmax$. One observes a smooth, oscillating excitation pattern, the first indication of which was already visible in Fig.~\ref{fig:F_2}. In particular, for $\Kmax = 0.186$, which is the lowest peak amplitude leading to maximum population transfer to the band $n = 2$, the final excited-band distribution is substantially narrower than the original one, while the distribution remaining in the lowest band is bimodal, corresponding to a wave packet moving in two opposite directions.

\begin{figure}[t]
\centering
\includegraphics[width = 0.8\linewidth]{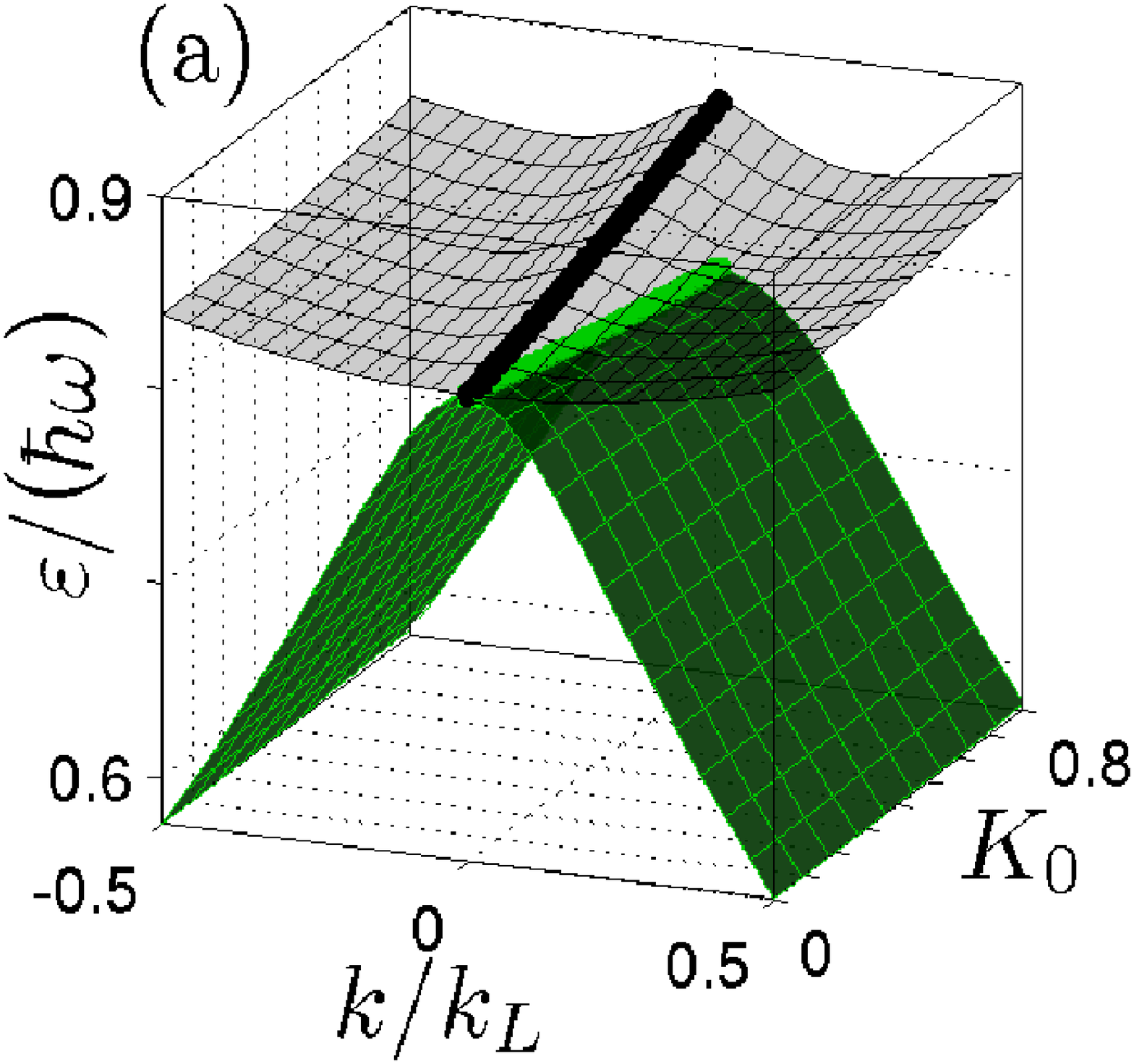}
\includegraphics[width = 0.8\linewidth]{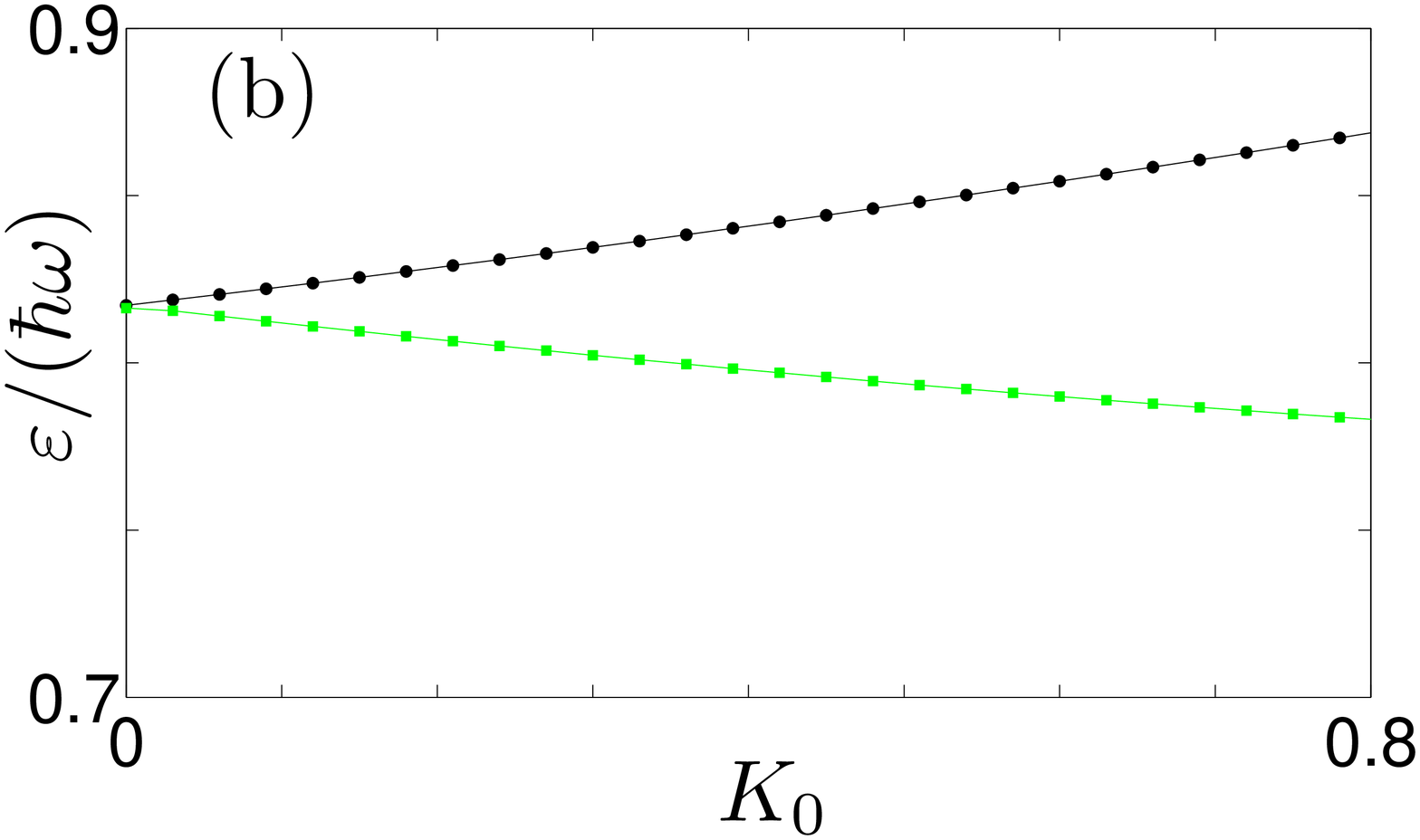}
\caption{(Color online) (a) Quasienergy surfaces underlying the excitation pattern observed in Fig.~\ref{fig:F_3}. The upper surface originates from the unperturbed Bloch band $n = 1$, the lower one from the band $n = 2$. Because of the resonant frequency, both surfaces are degenerate at $k/\kL = 0$ for vanishing instantaneous amplitude~$K_0$. The quasienergy lines at $k/\kL = 0$ are emphasized for better visibility. (b) Section through the surfaces at $k/\kL = 0$, showing	the removal of the initial degeneracy.}   
\label{fig:F_4}
\end{figure}

With the help of the tools assembled in the preceding section, these results can be understood in an almost intuitive manner, without the need to invoke much formalism. Figure~\ref{fig:F_4}(a) depicts the two quasienergy surfaces involved in the dynamics, emerging from the two unperturbed Bloch bands $n = 1$ and $n = 2$. Because of the Brillouin-zone structure of the quasienergy spectrum, the ``one-photon resonant'' driving frequency causes a degeneracy of both surfaces for vanishing instantaneous scaled amplitude $K_0 = \Fac a/(\hbar\omega)$. As a consequence, the initial wave packet is not placed on an individual quasienergy surface under the action of a pulse, but rather placed coherently on both surfaces: A Floquet expansion~\eqref{eq:FLR} yields contributions for both $n = 1$ and $n = 2$. Both parts of the wave function then react adiabatically to the slowly varying amplitude, each one picking up its own dynamical phase factors, given by the time integrals over the instantaneous quasienergies $\varepsilon_n^{\Fac}(k)$ encountered. When the driving amplitude goes to zero at the end of the pulse, both parts of the wave function produce an interference pattern which determines the final excitation probability: For each wave number $k$ sufficiently close to resonance, the transition probability to the first excited band is proportional to the expression~\cite{HolthausJust94}
\begin{equation}
	P_{1 \to 2}^{(k)} = 
	\sin^2\left(\frac{1}{2\hbar} \int_0^{\TP} \!\! \rd t \,	\left[ \varepsilon_1^{\Fac(t)}(k) - \varepsilon_2^{\Fac(t)}(k) \right] \right) \; .
\label{eq:RES}
\end{equation}   
As seen in Fig.~\ref{fig:F_4}(b), for $k/\kL = 0$ the quasienergy difference $\varepsilon_1^{\Fac}(0) - \varepsilon_2^{\Fac}(0)$ increases linearly with the driving amplitude, as is typical for a single-photon resonance~\cite{HolthausJust94}. Maximum excitation then is obtained when the argument of the squared sine in Eq.~\eqref{eq:RES} equals an odd-integer multiple of $\pi/2$, the first such maximum showing up for
\begin{equation}
	\frac{1}{\hbar} \int_0^{\TP} \!\! \rd t \, \left[ \varepsilon_1^{\Fac(t)}(k) - \varepsilon_2^{\Fac(t)}(k) \right] = \pm \pi \; .	
\label{eq:PIP}
\end{equation}
This is reminiscent of the familiar $\pi$-pulse condition; indeed, when the quasienergies are calculated analytically within the rotating-wave approximation, Eq.~\eqref{eq:PIP} reduces to the customary area theorem known from optical resonance~\cite{AllenEberly75,HolthausJust94}. But here we are confronted with the fact that this condition~\eqref{eq:PIP} can not be met simultaneously for all components $k$ with one single pulse shape: When it is satisfied for $k/\kL = 0$, the other components of the wave packet experience slightly or even strongly different quasienergies, depending on its initial width in $k$ space, as becomes evident when looking at Fig.~\ref{fig:F_4}(a). This is exactly what allows one to ``cut out'' a part of the momentum distribution, as was demonstrated in Fig.~\ref{fig:F_3}(c): Here the pulse shape is such that Eq.~\eqref{eq:PIP} indeed is satisfied for $k/\kL = 0$, leading to maximum transition probability in the center of the Brillouin zone. In contrast, the initial degeneracy at $k/\kL = 0$ has no effect on the wings of the initial distribution, so that these wings return adiabatically. As a result, the pulse transfers a relatively narrow central part of the initial distribution to the first excited Bloch band, leaving behind a symmetric bimodal distribution in the lowest one.        

\begin{figure}[t]
\centering
\includegraphics[width = 0.7\linewidth]{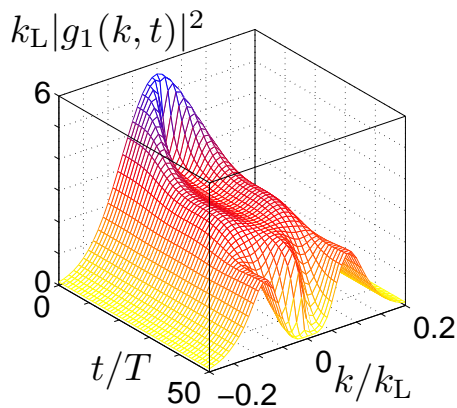}
\caption{(Color online) Floquet representation of the evolution of a wave packet initially prepared in the lowest Bloch band, under the action of a resonant $\pi$ pulse with $\Kmax = 0.186$. This figure shows how the final bimodal distribution $|g_1(k,\TP)|^2$ depicted in Fig.~\ref{fig:F_3}(c) appears after an initial reduction of the original density, corresponding to the partial occupation of the other resonantly coupled quasienergy surface; and after a period of almost adiabatic motion during the middle of the pulse.} 
\label{fig:F_5}
\end{figure}

It is then of particular interest to monitor the dynamics during such a pulse in the Floquet representation, instead of merely looking at the final distributions in the usual crystal-momentum representation. One such example, visualizing the the action of the very ``$\pi$ pulse'' considered above, is shown in Fig.~\ref{fig:F_5}. Observe how the evolution of $|g_1(k,t)|^2$ embodies the elements discussed before: The distribution soon is reduced to half its initial height, reflecting the occupation of the other quasienergy surface at the beginning of the pulse; then stays about constant during the pulses' middle part, reflecting approximately adiabatic motion; and develops the bimodal pattern only at its end, reflecting the final interference.

\subsection{Nonresonant, strong forcing}
\label{sec:S_3b}

\begin{figure}[t]
\centering
\includegraphics[width = 0.8\linewidth]{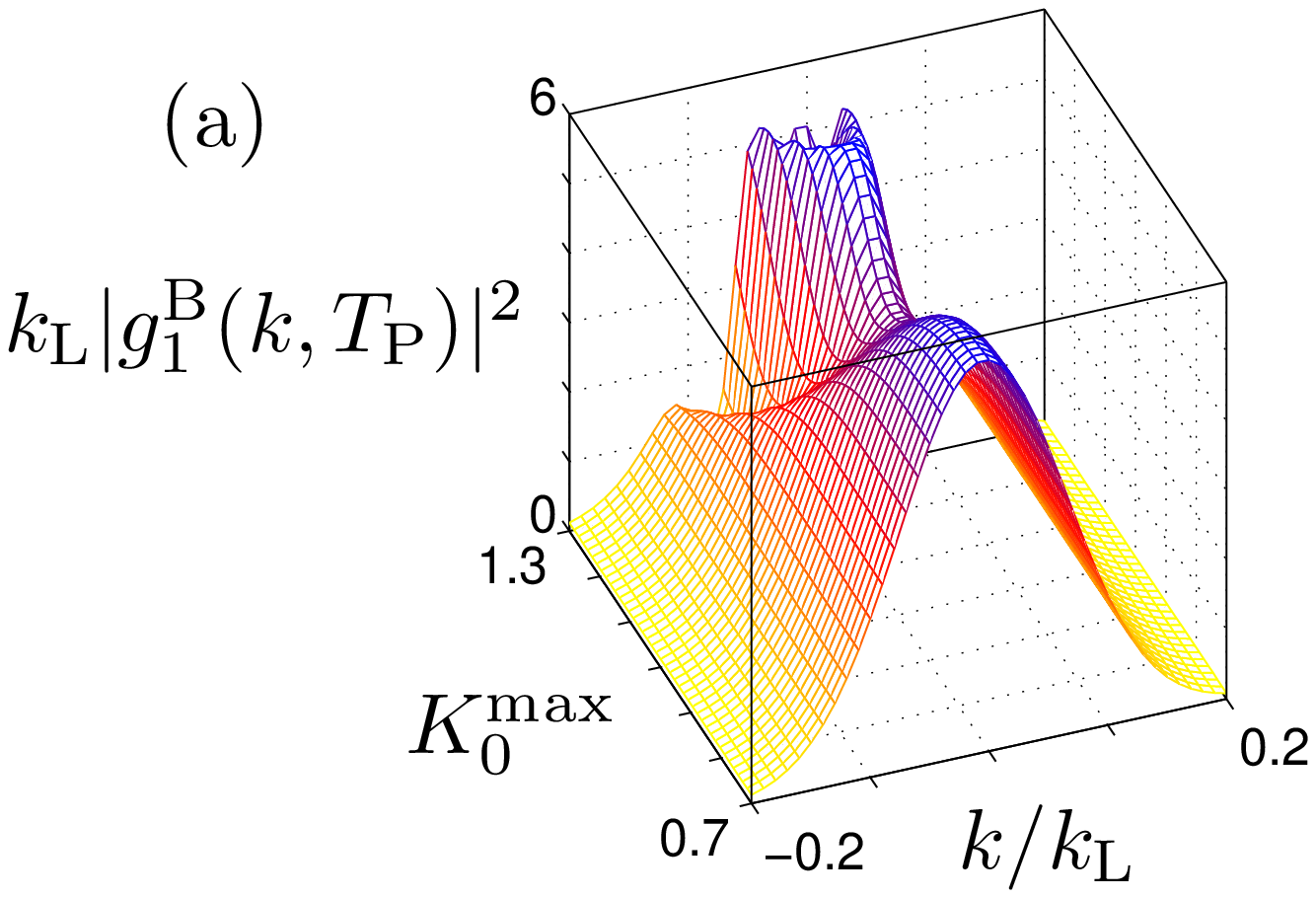}
\includegraphics[width = 0.8\linewidth]{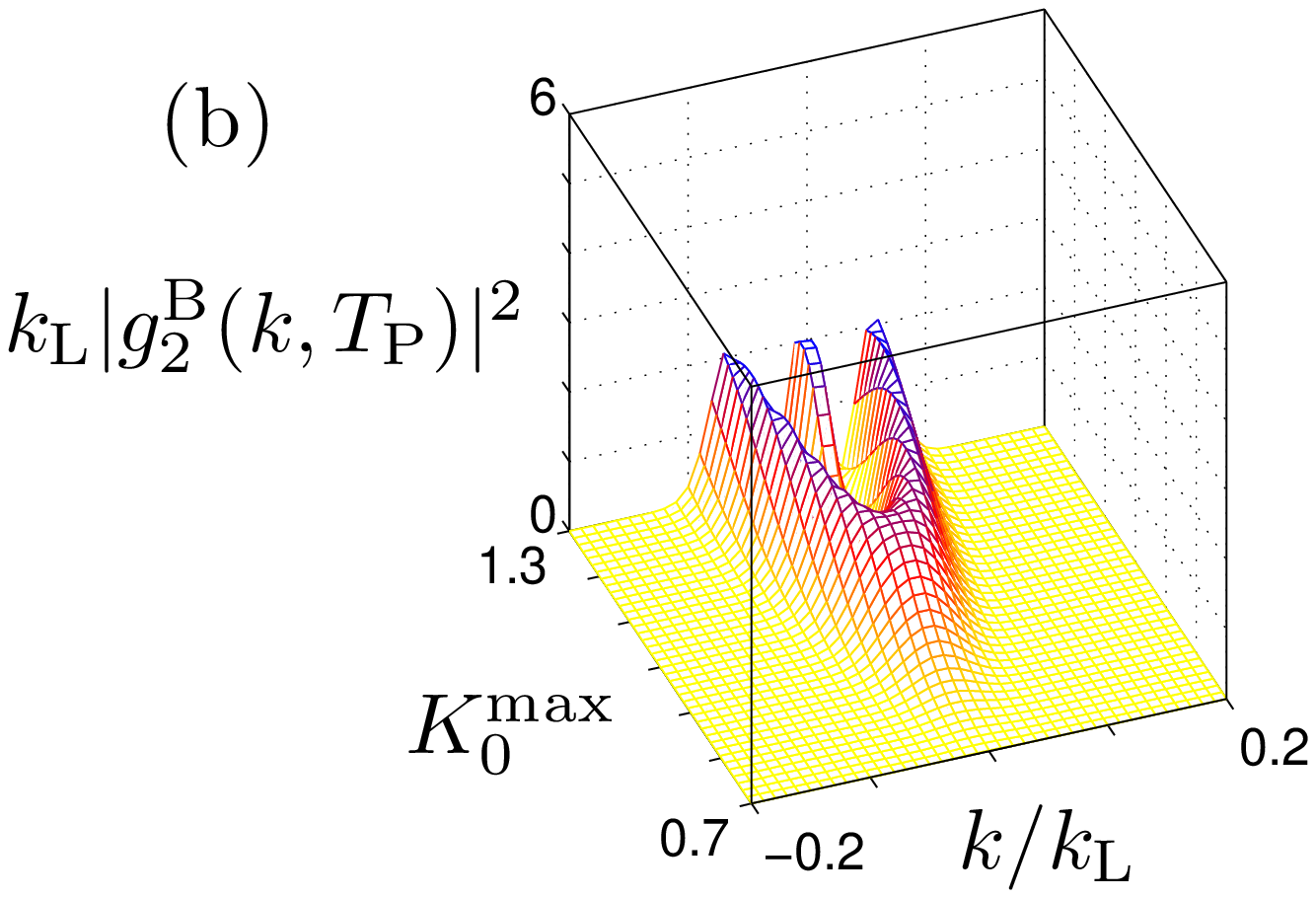}
\includegraphics[width = \linewidth]{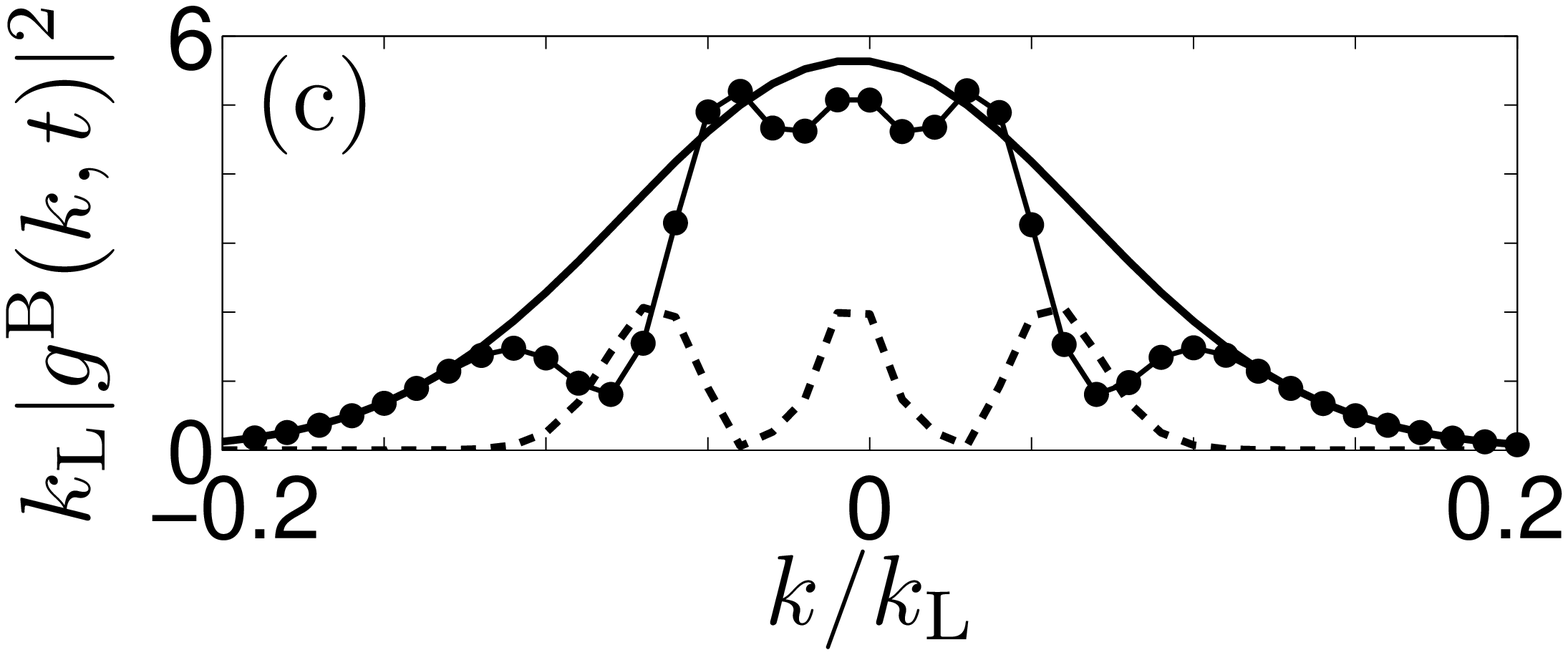}
\caption{(Color online) Final momentum distributions in the lowest (a) and in the first excited (b) Bloch band after the initial state has been exposed to pulses with squared-sine envelope~\eqref{eq:env} and length $\TP/T = 50$, with varying maximum scaled amplitudes $0.7 \le \Kmax \le 1.3$. Here the nonresonant scaled driving frequency is $\hbar\omega/\Er = 1.640$, as in Fig.~\ref{fig:F_1}. (c) compares the final distributions in the first (dotted) and in the second (dashed) bands to the initial distribution (full line) for $\Kmax = 1.3$.}
\label{fig:F_6}
\end{figure}

For the following second example of wave packet manipulation we again select the driving frequency $\omega = 1.640 \, \Er/\hbar$, as in the previous calculations having led to Fig.~\ref{fig:F_1}, but now we also consider pulses with larger scaled amplitudes~$\Kmax$. In Fig.~\ref{fig:F_6} we display the final distributions $|g_1^{\rm B}(k,\TP)|^2$ and $|g_2^{\rm B}(k,\TP)|^2$ as resulting from pulses with $0.7 \le \Kmax \le 1.3$. Even for $\Kmax = 0.8$ the initial distribution returns still undistorted, but for $\Kmax \approx 0.9$ strong interband transitions set in, leading to a trimodal excited-band distribution when $\Kmax = 1.3$.

\begin{figure}[t]
\centering
\includegraphics[width = 0.8\linewidth]{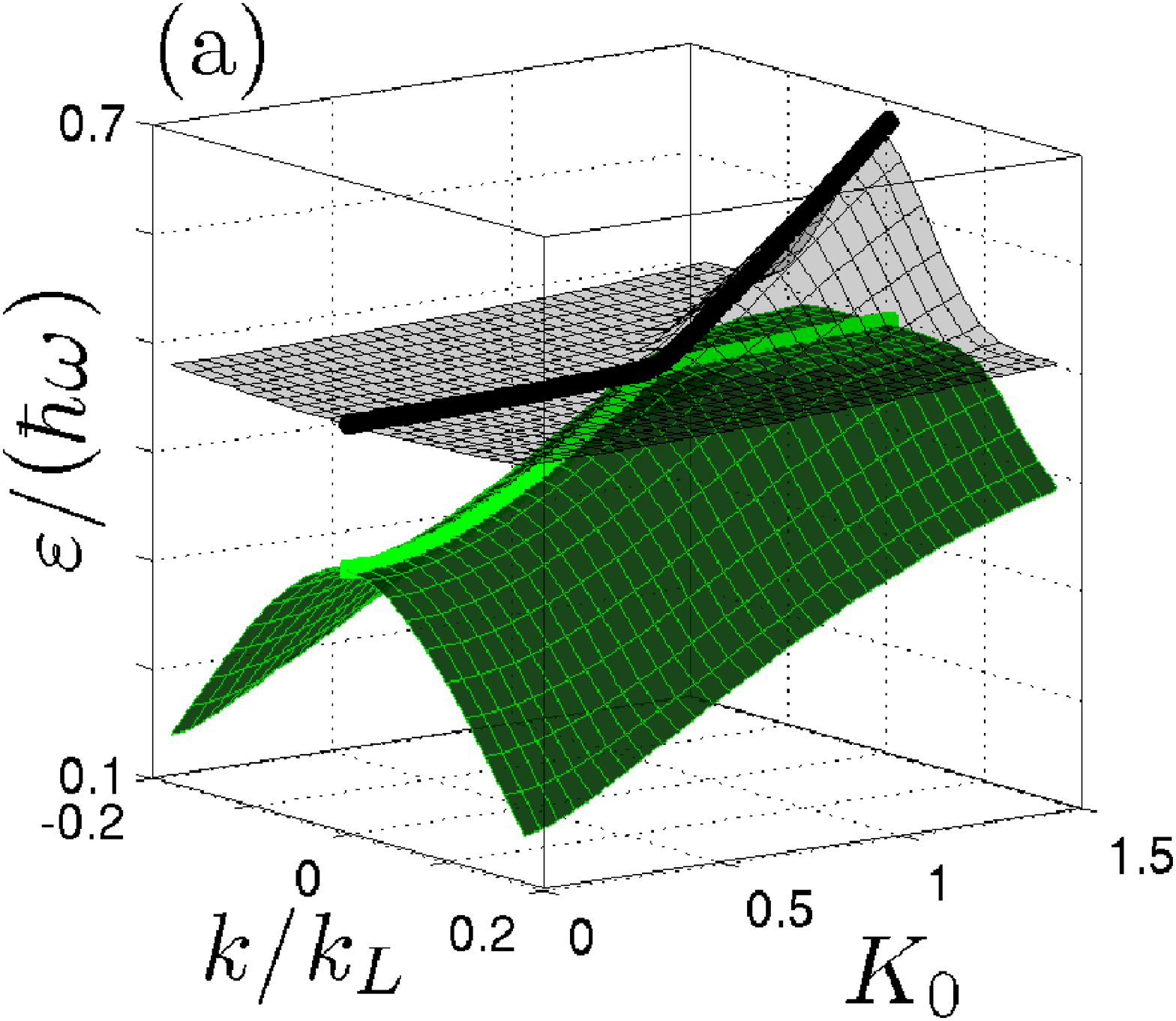}
\includegraphics[width = 0.8\linewidth]{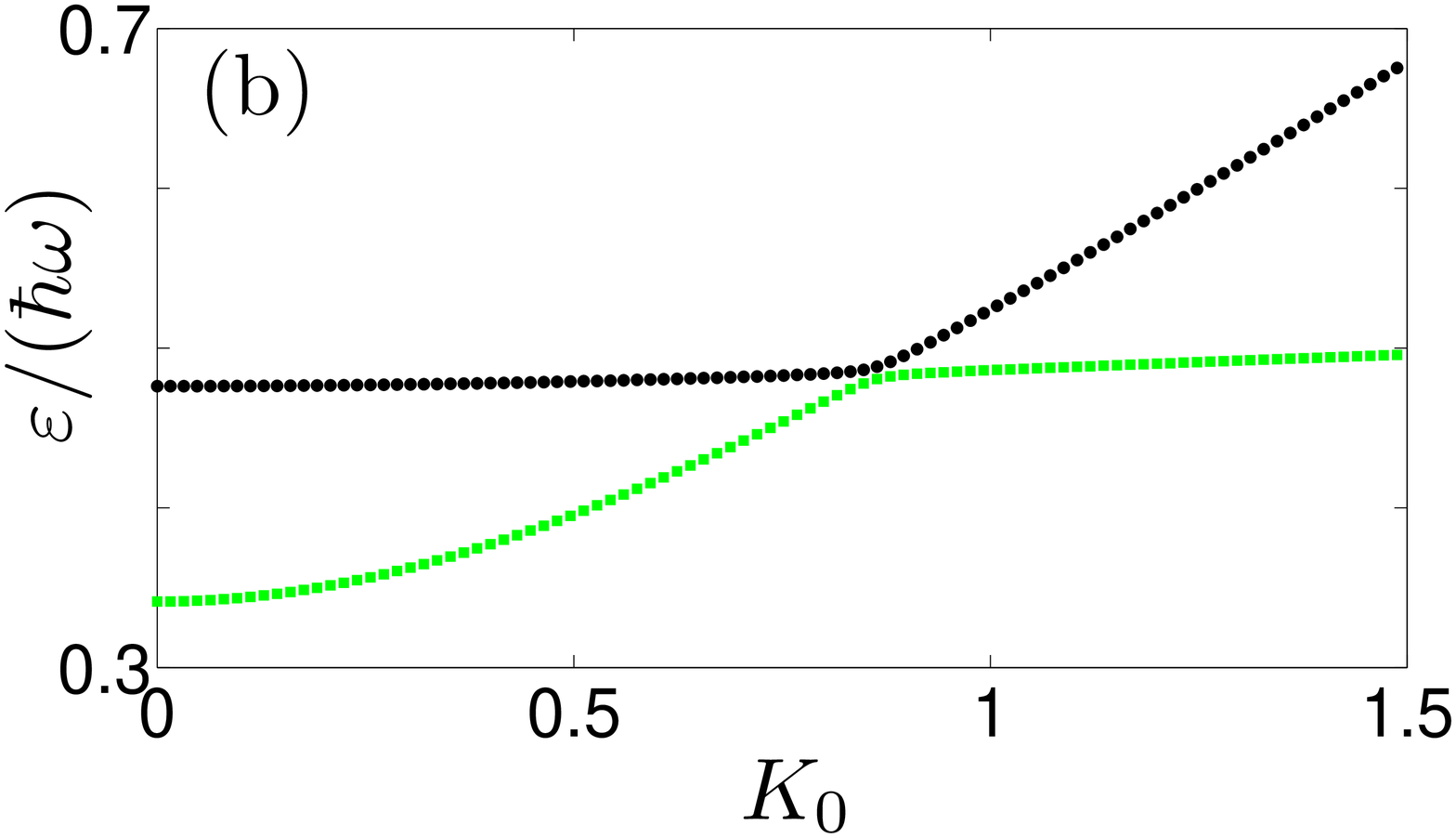}
\caption{(Color online) (a) Quasienergy surfaces underlying the excitation pattern observed in Fig.~\ref{fig:F_6}. The almost flat surface originates from the comparatively narrow Bloch band $n = 1$, the one with larger curvature from the band $n = 2$. Both surfaces undergo an avoided crossing along a parabola with apex at $k/\kL = 0$ and $K_0 \approx 0.9$. The quasienergy lines at $k/\kL = 0$ are emphasized for better visibility. (b) Section through the surfaces at $k/\kL = 0$, showing the narrow avoided crossing.}   
\label{fig:F_7}
\end{figure}

Once more the explanation for this response behavior is provided by the morphology of the quasienergy surfaces, shown in Fig.~\ref{fig:F_7}. The surface with comparatively low curvature originates from the Bloch band $n = 1$; this surface is penetrated by the one emerging from the Bloch band $n = 2$ along a parabola-shaped line with apex at $k/\kL = 0$ and $K_0 \approx 0.9$. Along this line the two surfaces exhibit a narrow avoided crossing. The quasienergy representatives plotted in Fig.~\ref{fig:F_7} are shifted against those continuously connected to the original energy bands by $+\hbar\omega$ ($n=1$) and by $-2\hbar\omega$ ($n=2$), respectively, so that the anticrossing marks a ``three-photon resonance.'' As long as the maximum pulse amplitude does not reach the apex of the anticrossing parabola, the momentum distribution merely moves adiabatically on its quasienergy surface, as already demonstrated in Fig.~\ref{fig:F_1}, and returns without notable modification. But when $\Kmax > 0.9$ the distribution has to pass the avoided-crossing line, resulting in partial Landau-Zener transitions to the other surface. Thus, the onset of excited-band population in the $\Kmax$-$k/\kL$ plane, as observed in Fig.~\ref{fig:F_6}(b), precisely reflects the locus of the band intersection.

Considering a pulse with $\Kmax > 0.9$, the different $k$ components thus encounter ``their'' respective avoided crossing at different amplitudes, and hence at different times. Therefore, the different components acquire quite different dynamical phase factors between their first passage through an avoided crossing during the rise of the pulse and the second passage taking place during the switch-off. Thus, for each $k$ one finds St\"uckelberg oscillations~\cite{Stueckelberg32} due to the interference of the parts having evolved on the two different surfaces, but their phases vary strongly within the Brillouin zone. This feature is the origin of the trimodal distribution $|g_2^{\rm B}(k,\TP)|^2$ shown in Fig.~\ref{fig:F_6}(c). As can be clearly seen in Fig.~\ref{fig:F_6}(b), the center lobe of this distribution already corresponds to the second St\"uckelberg maximum, whereas the two outer lobes still are associated with the first. 
 
\begin{figure}[t]
\centering
\includegraphics[width = 0.7\linewidth]{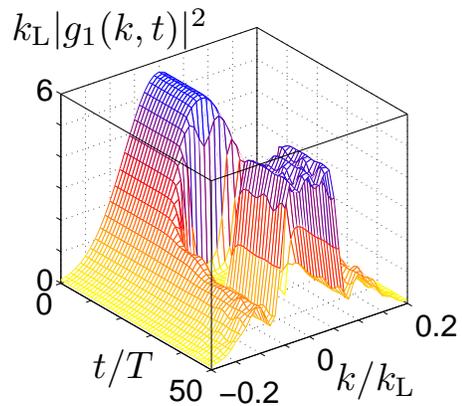}
\caption{(Color online) Floquet representation of the evolution of a wave packet initially prepared in the lowest Bloch band, under the action of a nonresonant pulse with $\Kmax = 1.3$. After an initial period of almost adiabatic motion, the center of the distribution undergoes partial Landau-Zener transitions to the anticrossing surface	depicted in Fig.~\ref{fig:F_7}, and later returns to the initial surface, subject to St\"uckelberg oscillations.} 
\label{fig:F_8}
\end{figure}

In Fig.~\ref{fig:F_8} we show the evolution of the Floquet distribution $|g_1(k,t)|^2$ during the pulse with maximum amplitude $\Kmax = 1.3$. Evidently the outer wings of this distribution move adiabatically, not encountering the avoided-crossing line, whereas the central part jumps to the anticrossing surface with high probability when passing this line, returning later when the line is hit a second time. 

We remark that by suitable choices of the frequency one may likewise design pulses which involve higher quasienergy bands: If the Bloch band $n = 1$ is slightly detuned from a multiphoton resonance with a higher band, and if the ac Stark shift forces the quasienergy surfaces emerging from these two bands into an anticrossing, one can exploit the corresponding multiphotonlike Landau-Zener transitions in the same manner as in the example considered here.

\section{Phase effects}
\label{sec:S_4}

\begin{figure}[t]
\centering
\includegraphics[width = 0.7\linewidth]{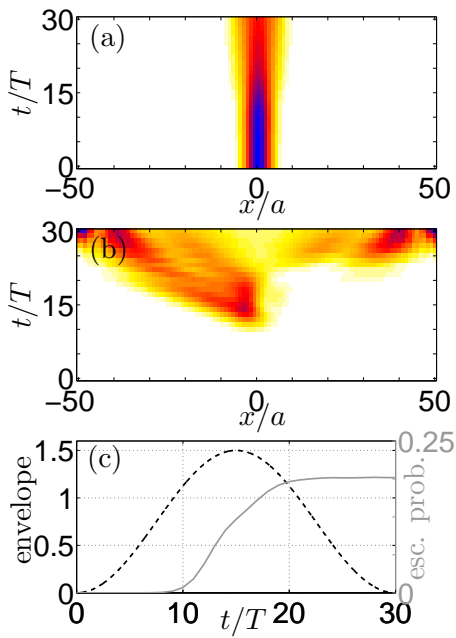}
\caption{(Color online) Contour plot of the real-space density $|\psi_1(x,t)|^2$ associated with the lowest Bloch energy band (a), and of the density $|\psi_2(x,t)|^2$ associated with the first excited band (b), as resulting from a pulse with $\hbar\omega/\Er = 1.640$, $\Kmax = 1.5$, and $T/\TP = 30$. The phase $\varphi$ in Eq.~\eqref{eq:PUP} has been set to zero. (c) shows the gradual loss of population from the lowest band.}
\label{fig:F_9}
\end{figure}

The solutions to the quasienergy eigenvalue equation~\eqref{eq:EVE} refer to perfectly periodic driving, and thus do not change when the force $F(t) = \Fac \sin(\omega t)$ is replaced by $F(t) = \Fac \sin(\omega t + \varphi)$, except for a trivial shift of the time coordinate. Accordingly, as long as the pulse dynamics are fully adiabatic they are not affected by the phase~$\varphi$. This is different, however, under non-adiabatic conditions. Then the transitions effectuated by a pulse may strongly depend on $\varphi$, so that this phase offers an additional handle of control. For demonstration, we replace the previous pulses~\eqref{eq:PUL} by  
\begin{equation}
	F(t) = \Fac^{\max} s(t) \sin(\omega t + \varphi) \; ,
\label{eq:PUP}	
\end{equation}
maintaining the interval $0 \le t \le \TP$ as the active pulse period and employing the same squared-sine envelope~\eqref{eq:env} as before. Having set $\omega = 1.640 \, \Er/\hbar$, $\Kmax = 1.5$, and $T = 30 \, \TP$, Fig.~\ref{fig:F_9} shows contour plots of the densities $|\psi_1(x,t)|^2$ and $|\psi_2(x,t)|^2$ associated with the lowest two Bloch bands, together with the transition dynamics, for $\varphi = 0$. Evidently the transfer of probability to the excited band does not proceed symmetrically in space.  
 
\begin{figure}[t]
\centering
\includegraphics[width = 0.7\linewidth]{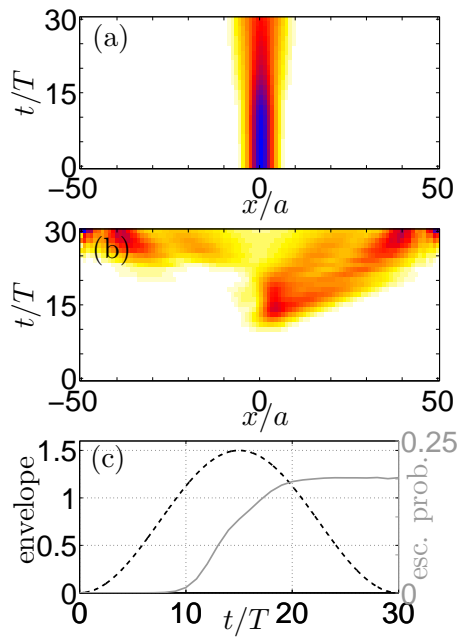}
\caption{(Color online) As Fig.~\ref{fig:F_9}, but for $\varphi = \pi$.}
\label{fig:F_10}
\end{figure}

Figure~\ref{fig:F_10} then depicts the corresponding results for $\varphi = \pi$. While the distribution $|\psi_2(x,t)|^2$ here is the exact mirror image of that displayed in Fig.~\ref{fig:F_9}(b), the depopulation of the lowest band proceeds in exactly the same manner as before. This finding is easily explained: Replacing $\varphi$ by $\varphi + \pi$, and simultaneously replacing $x$ by $-x$, leaves the interaction term $-F(t)x$ in the Hamiltonian~\eqref{eq:tildeH} invariant, so that a phase shift by $\pi$ is equivalent to spatial inversion. Needless to say, one can also select other values of $\varphi$ and produce results not predictable by simple symmetry considerations. For example, the outcome of a passage through an avoided crossing may depend on the particular instantaneous phase at which this anticrossing is met. More generally, the possible effects of the phase of the carrier frequency with respect to the pulse envelope closely resemble corresponding effects encountered in laser-atom interaction~\cite{PaulusEtAl01}.

\section{Conclusions and further visions}
\label{sec:S_5}

In this paper we have discussed mechanisms which govern the response of a single particle in a cosine lattice to pulsed homogeneous forcing, with a view toward controlling weakly interacting Bose-Einstein condensates in shaken optical lattices. While several successful experiments with ultracold atoms in strongly ac-forced optical lattices have already been reported~\cite{LignierEtAl07,EckardtEtAl09,SiasEtAl08,ZenesiniEtAl09,StruckEtAl11,ChenEtAl11,AlbertiEtAl09,HallerEtAl10}, the systematic exploration of the possibilities of coherent control opened up, e.g., by deliberate pulse shaping, is likely to break new ground in ultracold-atom physics.

This optimistic view is suggested by a simple parallel: Atomic and molecular physics in itself, not invoking the use of lasers, already is a fascinating and important subject, but it becomes infinitely more rich when lasers come into play, allowing one, on the one hand, to do precision spectroscopy, and to perform deliberate state manipulations on the other. By the same token, ultracold atoms in optical lattices offer access to much fundamental physics, but there is a host of further options when applying inertial forces, either in the form of an ac~drive for creating dressed matter waves, or in the form of carefully designed pulses in order to exert active coherent control.

Here we have focused on elementary mechanisms of control deriving from pulses with a smooth envelope, leading to adiabatic motion of a wave packet's momentum distribution on quasienergy surfaces created by spatiotemporal Bloch waves and to deviations from adiabaticity which can be purposefully exploited for reaching target states which may not be accessible by other means. The accompanying Floquet picture offers the distinct advantage that it enables one to adapt many concepts developed for the theoretical description of laser-matter interaction, such as the notion of $\pi$ pulses. In our opinion, the actual implementation and observation of such $\pi$ pulses with dilute Bose-Einstein condensates in shaken optical lattices constitutes an {\em experimentum crucis\/}: If this can be done, many further related control scenarios will be equally viable, also involving variations of the driving frequency. 

As a future perspective it seems particularly rewarding to also carry over advanced strategies devised for controlling molecular dynamics and even chemical reactions by specifically engineered laser pulses~\cite{JudsonRabitz92,Rice92,AssionEtAl98,ShapiroBrumer03,Baumert11}. Such techniques often involve feedback loops for optimizing the pulse characteristics with the help of genetic learning algorithms; this approach is tantamount to ``teaching lasers to control molecules''~\cite{JudsonRabitz92}. When working with Bose-Einstein condensates in optical lattices, control pulses can be applied by piezo-electrically juggling the position of a mirror which reflects the lattice-generating laser beam back into itself, thus giving rise to a moving standing light wave in the laboratory frame, and to a corresponding inertial force in the comoving frame of reference. By sheer analogy, one then would ``teach mirrors to control condensates.'' In this manner one could assess the ``reachability problem'' for condensates in optical lattices and explore whether one can reach any preselected final-state distribution with the help of a proper pulse sequence. In particular, it would be of major interest to apply such strategies for creating mesoscopic Schr\"odinger-cat-like states, that is, quantum superposition states of condensates.

\begin{acknowledgments}
This work was supported by the Deutsche Forschungsgemeinschaft 
under Grant No.~Ho~1771/6. 
\end{acknowledgments}


\begin{thebibliography}{99}

\bibitem{GreinerEtAl02} M. Greiner, O. Mandel, T. Esslinger, T. W. H\"ansch,
	and I. Bloch,
	Nature (London) {\bf 415}, 39 (2002).

\bibitem{MorschOberthaler06} O. Morsch and M. Oberthaler,
    	Rev. Mod. Phys. {\bf 78}, 179 (2006).
    
\bibitem{LewensteinEtAl07} M. Lewenstein, A. Sanpera, V. Ahufinger, 
	B. Damski, A. Sen, and U. Sen,
	Adv. Phys. {\bf 56}, 243 (2007).
    
\bibitem{BlochEtAl08} I. Bloch, J. Dalibard, and W. Zwerger,
	Rev. Mod. Phys. {\bf 80}, 885 (2008).

\bibitem{DreseHolthaus97} K. Drese and M. Holthaus,
	Phys. Rev. Lett. {\bf 78}, 2932 (1997).

\bibitem{LignierEtAl07} H. Lignier, C. Sias, D. Ciampini, Y. Singh,
    	A. Zenesini, O. Morsch, and E. Arimondo,
    	Phys. Rev. Lett. {\bf 99}, 220403 (2007).
    
\bibitem{EckardtEtAl09} A. Eckardt, M. Holthaus, H. Lignier, A. Zenesini,
	D. Ciampini, O. Morsch, and E. Arimondo,
	Phys. Rev. A {\bf 79}, 013611 (2009).
   
\bibitem{SiasEtAl08} C. Sias, H. Lignier, Y. P. Singh, A. Zenesini,
    	D. Ciampini, O. Morsch, and E. Arimondo,
    	Phys. Rev. Lett. {\bf 100}, 040404 (2008).

\bibitem{ZenesiniEtAl09} A. Zenesini, H. Lignier, D. Ciampini, O. Morsch,
    	and E. Arimondo, 
	Phys. Rev. Lett. {\bf 102}, 100403 (2009).
		
\bibitem{EckardtEtAl05} A. Eckardt, C. Weiss, and M. Holthaus,
    	Phys. Rev. Lett. {\bf 95}, 260404 (2005).
	
\bibitem{StruckEtAl11} 	J. Struck, C. \"Olschl\"ager, R. Le Targat, 
	P. Soltan-Panahi, A. Eckardt, M. Lewenstein, P. Windpassinger, 
	and K. Seng\-stock,
	Science {\bf 333}, 996 (2011). 
	
\bibitem{ChenEtAl11} Y.-A. Chen, S. Nascimb\`{e}ne, M. Aidelsburger, 
	M. Atala, S. Trotzky, and I. Bloch,
	Phys. Rev. Lett. {\bf 107}, 210405 (2011).

\bibitem{KudoEtAl09} K. Kudo, T. Boness, and T. S. Monteiro,
	Phys. Rev. A {\bf 80}, 063409 (2009).									 	  	
	
\bibitem{TokunoGiamarchi11} A. Tokuno and T. Giamarchi,
	Phys. Rev. Lett. {\bf 106}, 205301 (2011).
	
\bibitem{TsujiEtAl11} N. Tsuji, T. Oka, P. Werner, and H. Aoki,
	Phys. Rev. Lett. {\bf 106}, 236401 (2011).
	
\bibitem{Holthaus01} M. Holthaus,
        Phys. Rev. A {\bf 64}, 011601 (2001).	
	
\bibitem{PolettiKollath11} D. Poletti and C. Kollath,
	Phys. Rev. A {\bf 84}, 013615 (2011). 	 	

\bibitem{IvanovEtAl08} V. V. Ivanov, A. Alberti, M. Schioppo, G. Ferrari, 
	M. Artoni, M. L. Chiofalo, and G. M. Tino,
	Phys. Rev. Lett. {\bf 100}, 043602 (2008).
	
\bibitem{GustavssonEtAl08} M. Gustavsson, E. Haller, M. J. Mark, J. G. Danzl, 
	G. Rojas-Kopeinig, and H.-C. N\"agerl,
   	Phys. Rev. Lett. {\bf 100}, 080404 (2008).
	
\bibitem{AlbertiEtAl09} A. Alberti, V. V. Ivanov, G. M. Tino, and G. Ferrari,
	Nat. Phys. {\bf 5}, 547 (2009).
	
\bibitem{HallerEtAl10} E. Haller, R. Hart, M. J. Mark, J. G. Danzl,
	L. Reich\-s\"ollner, and H.-C. N\"agerl,
	Phys. Rev. Lett. {\bf 104}, 200403 (2010).

\bibitem{ArlinghausHolthaus10} S. Arlinghaus and M. Holthaus,
	Phys. Rev. A {\bf 81}, 063612 (2010).

\bibitem{ArlinghausHolthaus11} S. Arlinghaus and M. Holthaus,
	Phys. Rev. B {\bf 84}, 054301 (2011).	

\bibitem{AllenEberly75} L. Allen and J. H. Eberly, 
	{\em Optical Resonance and Two-Level Atoms\/} 
	(Dover, New York, 1987).

\bibitem{JudsonRabitz92} R. S. Judson and H. Rabitz,
	Phys. Rev. Lett {\bf 68}, 1500 (1992).

\bibitem{Rice92} S. A. Rice,
	Science {\bf 258}, 412 (1992).
	
\bibitem{AssionEtAl98} A. Assion, T. Baumert, M. Bergt, T. Brixner, B. Kiefer,
	V. Seyfried, M. Strehle, and G. Gerber,
	Science {\bf 282}, 919 (1998).
	
\bibitem{ShapiroBrumer03} M. Shapiro and P. Brumer,
	Rep. Prog. Phys. {\bf 66}, 859 (2003).	
	
\bibitem{Baumert11} T. Baumert,
	Nat. Phys. {\bf 7}, 373 (2011).

\bibitem{Graham92} R. Graham, M. Schlautmann, and P. Zoller,
	Phys. Rev. A {\bf 45}, 	R19 (1992).

\bibitem{Callaway76} J. Callaway,
	{\em Quantum Theory of the Solid State\/}
	(Academic Press, New York, 1976).

\bibitem{AshcroftMermin76} N. W. Ashcroft and N. D. Mermin,
	{\em Solid State Physics\/}
	(Harcourt, Fort Worth, 1976), Chap.~12.

\bibitem{Kittel87} C. Kittel,
	{\em Quantum Theory of Solids}, 2nd rev.\ ed.\
	(Wiley, New York, 1987).

\bibitem{Floquet83} G. Floquet,
	Ann. Scit. Ec. Normale Super. {\bf 12}, 47 (1883).
	
\bibitem{Eastham73} M. S. P. Eastham, 
	{\em The Spectral Theory of Periodic Differential Equations\/}, 
	Texts in Mathematics (Scottish Academic Press, Edinburgh, 1973).
	
\bibitem{Kuchment93} P. Kuchment, 
	{\em Floquet Theory for Partial Differential Equations\/}
	(Birkh\"auser, Basel, 1993).
			
\bibitem{Zeldovich66} Ya. B. Zel'dovich,
	Zh. Eksp. Teor. Fiz. {\bf 51}, 1492 (1966)
	[Sov. Phys. JETP {\bf 24}, 1006 (1967)].	  	

\bibitem{ChuTelnov04} S. I. Chu and D. A. Telnov,
	Phys. Rep. {\bf 390}, 1 (2004). 

\bibitem{Sambe73} H. Sambe,
	Phys. Rev. A  {\bf 7}, 2203 (1973).

\bibitem{DreseHolthaus99} K. Drese and M. Holthaus,
	Eur. Phys. J. D {\bf 5}, 119 (1999).

\bibitem{BornOppenheimer27} M. Born and R. Oppenheimer, 
	Ann. Physik {\bf 389}, 457 (1927). 
		
\bibitem{BornFock28} M. Born and V. Fock,
	Z. Phys. {\bf 51}, 165 (1928).

\bibitem{Berry84} M. V. Berry,
	Proc. R. Soc. London, Ser. A {\bf 392}, 45 (1984).

\bibitem{HolthausJust94} M. Holthaus and B. Just,
	Phys. Rev. A {\bf 49}, 1950 (1994).

\bibitem{Stueckelberg32} E. C. G. St\"uckelberg,
	Helv. Phys. Acta {\bf 5}, 369 (1932).
	
\bibitem{PaulusEtAl01} G. G. Paulus, F. Grasbon, H. Walther, P. Villoresi,
	M. Nisoli, S. Stagira, E. Priori, and S. De Silvestri,
	Nature (London) {\bf 414}, 182 (2001).	
						
\end{thebibliography}
\end{document}